\documentstyle[twoside, epsf]{article}

\input ibvs2.sty

\begin{document}
\IBVShead{5868}{00 Month 200x}

\IBVStitle{Two pairs of interacting EBs towards the LMC in the OGLE database}

\IBVSauth{Ofir, A.$^1$}

\IBVSinst{School of Physics and Astronomy, Raymond and Beverly Sackler Faculty of Exact Sciences, Tel Aviv University, Tel Aviv, Israel}

\SIMBADobjAlias{}
\IBVStyp{}
\IBVSkey{}
\IBVSabs{This is the second}
\IBVSabs{And this is the last one.}

\begintext
Manual browsing through the online OGLE LMC database\footnote{http://ogle.astrouw.edu.pl/} [Wyrzykowski \textit{et al.} 2003] revealed that eclipsing binary OGLE 051343.14-691837.1 ($P_1$=3.57798d) is significantly more noisy than other stars with similar brightness, and indeed another EB was subsequently found in it's residuals ($P_2$=5.36655d). This second EB has a period of almost exactly 1.5 times the period of the first EB. To better disentangle the two signals we simultaneously fitted a truncated Fourier series with N terms (N=22) for each period, plus a zero-point term - for a total of 45 fitted coefficients. We then rejected outliers and repeated untill convergence. This allowed us to better visualize both EBs (see attached figure) - and it is easy to see that the two signals are not different harmonics of the same system but rather two distinct EBs. Since OGLE's telescope PSF is rather small, and because of the apparent resonance between the two binaries, we believe it is highly unlikely that this is chance alignment, and that the more probable explanation is of a rather compact hierarchical system of two pairs of EBs in 3:2 resonance. Interestingly, it seems that all 4
stars are rather massive as both EBs show very significant ellipsoidal
variation. With this scenario in mind, the fact that both pairs of stars are EBs means that some degree of co-planarity also exist - further supporting the interacting-pairs hypothesis.

\IBVSfigNoCaption{9cm}{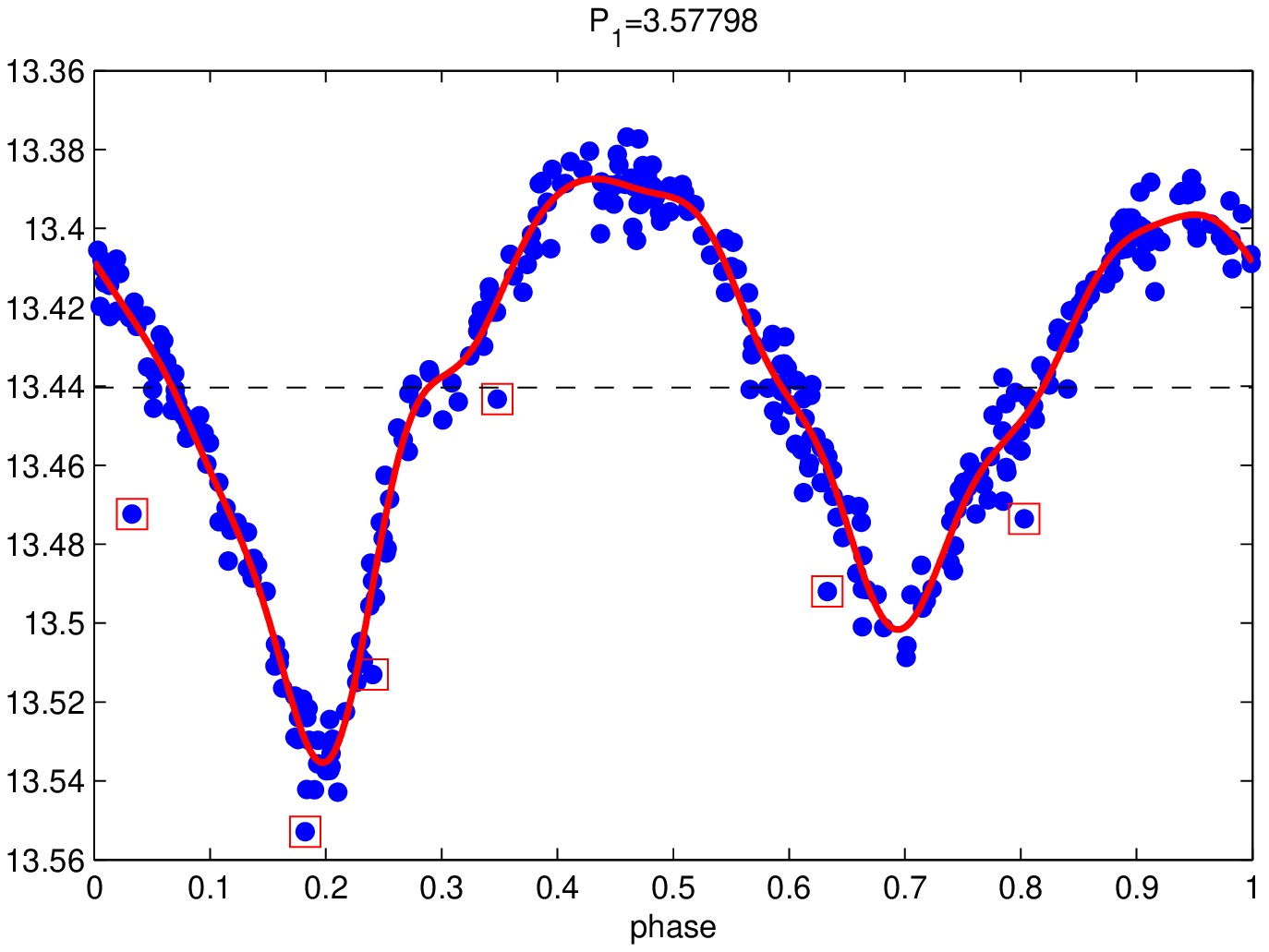} 
\IBVSfig{9cm}{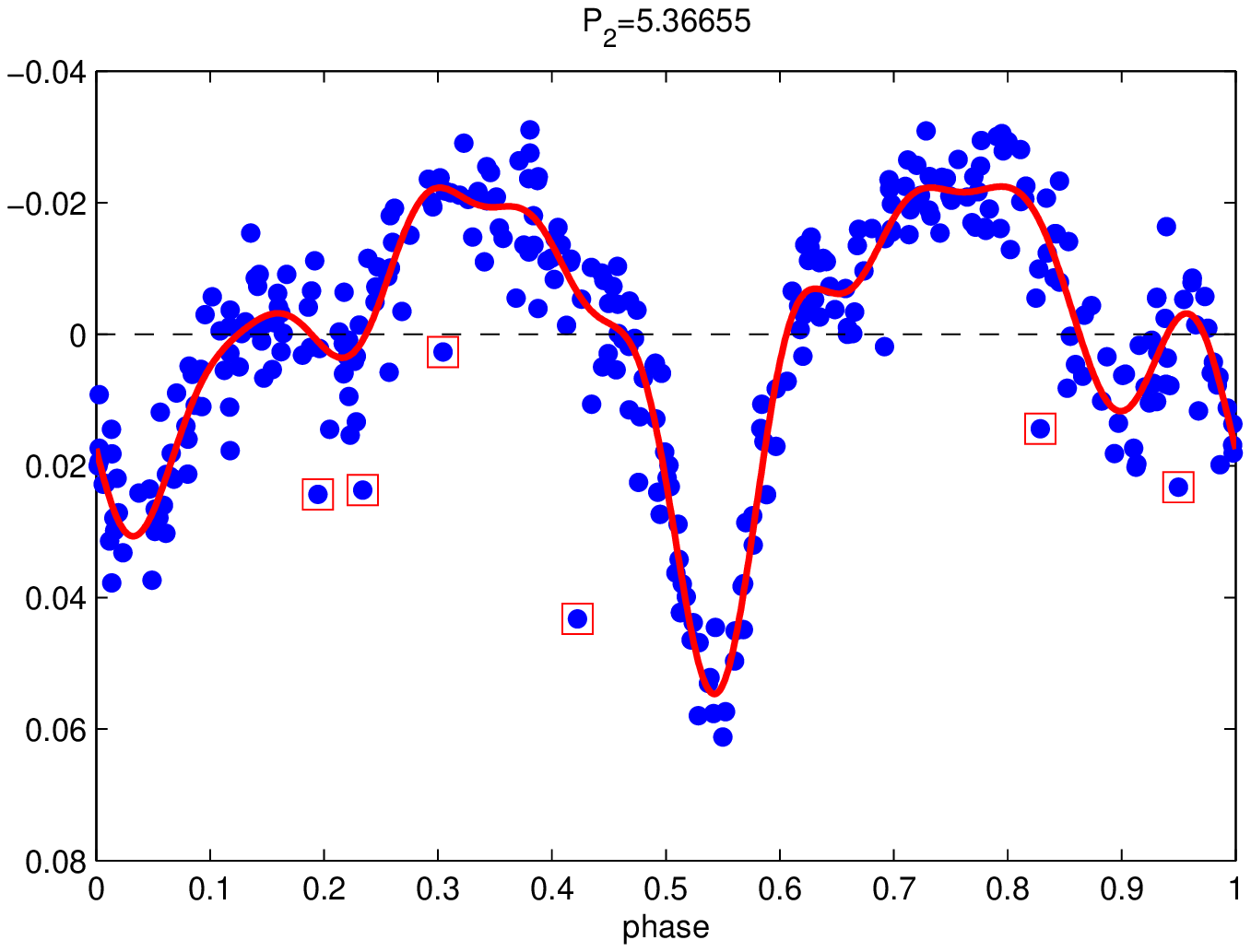}{
The red solid lines are smoothed light curves using a truncated Fourier series 
- see text for details (for each EB the other EB signal is removed). Outliers are marked in red boxes.}

\references

Wyrzykowski \textit{et al.} (2003), Acta Astron., 53, 1.

\endreferences

\end{document}